\documentclass[3p, times]{elsarticle}
\usepackage{graphicx}
\usepackage{color}
\usepackage{amssymb}
\usepackage{amsmath}

\usepackage [english]{babel}
\usepackage [autostyle, english = american]{csquotes}
\MakeOuterQuote{"}

\newcommand{\eps}{\varepsilon}

\def\l{\lambda}

\def\pr{{\rm prob}}

\def\beq{\begin{equation}}
\def\ee{\end{equation}}
\def\bi{\begin {itemize}}
\def\ei{\end{itemize}}

\def\eps{\epsilon}

\def\lsim
{\protect \raisebox{-0.75ex}[-1.5ex]{$\;\stackrel{<}{\sim}\;$}}
\def\gsim
{\protect \raisebox{-0.75ex}[-1.5ex]{$\;\stackrel{>}{\sim}\;$}}
\def\lsimeq
{\protect \raisebox{-0.75ex}[-1.5ex]{$\;\stackrel{<}{\simeq}\;$}}
\def\gsimeq
{\protect \raisebox{-0.75ex}[-1.5ex]{$\;\stackrel{>}{\simeq}\;$}}

\def\l{\lambda}

\def\pr{{\rm prob}}

\def\t{_{\rm tot}}

\def\njp{{n_j}^+}
\def\njm{{n_j}^-}

\def\c{{\cal C}}

\def\tot{_{\rm tot}}

\def\beq{\begin{equation}}
\def\ee{\end{equation}}

\def\bi{\begin {itemize}}
\def\ei{\end{itemize}}

\def\eps{\epsilon}

\def\wmn{w_{mn}}
\def\wnm{w_{nm}}

\def\dr{^{\rm dr}}
\def\ex{^{\rm ex}}
\def\hk{^{\rm hk}}
\def\sys{^{\rm sys}}

\def\I{{\cal I}}

\def\wout{w^{\rm out}}
\def\wcont{w^{\rm cont}}

\def\Wout{W^{\rm out}}
\def\Wcont{W^{\rm cont}}

\def\cont{^{\rm cont}}
\def\out{^{\rm out}}
\def\i{_{i}}
\def\iii{_{iii}}
\def\nn{\nonumber}

\begin{document}

\title{Stochastic thermodynamics: From principles to the cost of precision}
\author{ Udo Seifert }
\ead{useifert@theo2.physik.uni-stuttgart.de}

\address{ {II.} Institut f\"ur Theoretische Physik, Universit\"at Stuttgart,
  70550 Stuttgart, Germany\\~\\ }

\begin{abstract}

~\\~
In these lecture notes, the basic principles of stochastic thermodynamics are developed
starting with a closed system in contact with a heat bath. A trajectory undergoes Markovian
transitions between observable meso-states that correspond to a coarse-grained
description of, e.g., a biomolecule or a biochemical network. By separating the
closed system into a core system and into reservoirs for ligands and reactants  that bind
to, and react with the core system, a description as an open system controlled by chemical potentials and possibly an external force 
is achieved. Entropy production and further thermodynamic quantities  defined along a trajectory obey various fluctuation
theorems. For describing fluctuations in a non-equilibrium steady state
in the long-time limit, the concept of a rate function for large deviations from the mean behaviour
is derived from the weight of a trajectory. Universal bounds on this rate function follow
which prove and generalize the
thermodynamic uncertainty relation that quantifies the inevitable trade-off between
cost and precision of any biomolecular process. Specific illustrations are given for
molecular motors, Brownian clocks and enzymatic networks that show how these tools can
be used for thermodynamic inference of hidden properties of a system.


\end{abstract}

\begin{keyword}
\end{keyword}

\maketitle

\def\lsim
{\protect \raisebox{-0.75ex}[-1.5ex]{$\;\stackrel{<}{\sim}\;$}}

\def\gsim
{\protect \raisebox{-0.75ex}[-1.5ex]{$\;\stackrel{>}{\sim}\;$}}

\def\lsimeq
{\protect \raisebox{-0.75ex}[-1.5ex]{$\;\stackrel{<}{\simeq}\;$}}

\def\gsimeq
{\protect \raisebox{-0.75ex}[-1.5ex]{$\;\stackrel{>}{\simeq}\;$}}

\def\vs{\vskip 1cm}

\def\Ek{E_{\rm {k}}}
\def\Ec{E_{\rm c}}
\def\Ech{\hat E_{\rm c}}
\def\Ekh{{{\hat E}_{\rm k}}}
\def\Ekhs{\partial_E \hat E_{\rm k}}
\def\Eh{\hat E}
\def\Ok{\Omega_{\rm k}}
\def\Oc{\Omega_{\rm c}}
\def\O{\Omega}
\def\bk{\beta_{\rm k}}
\def\bc{\beta_{\rm c}}
\def\bks{\partial_E \beta_{\rm k}}
\def\bcs{\partial_E \beta_{\rm c}}

\def\bS{\bar S}
\def\S{{\cal S}}

\def\k{_{\rm k}}
\def\c{_{\rm c}}

\def\dek{(\Delta \Ek)^2}

\def\bb{\bar \beta}

\def\ph{p^{\rm hist}}
\def\t{{\cal T}}

\def\pb{{\bf p}}
\def\qb{{\bf q}}

\def\tt{_{\rm tot}}
\def\sy{_{\rm sys}}

\def\xib{{\boldsymbol \xi}}
\def\H{{\cal H}}
\def\P{{\cal P}}

\def\tO{\tilde\Omega}
\def\Og{\tilde \Omega_G}

\def\eff{_{\rm eff}}
\def\mic{_{\rm mic}}
\def\can{_{\rm can}}
\def\gc{_{\rm grc}}
\def\tZ{\tilde Z}
\def\U{{\cal U}}

\def\hnull{h}



\def\kij{K_{IJ}}
\def\kji{K_{JI}}
\def\Dij{\Delta_{IJ}}
\def\dij{\Delta_{ij}}
\def\pa{p_\xi}
\def\pxi{P(\xi|I)}
\def\xai{{\xi\in I}}
\def\eq{^{e}}
\def\ss{^{s}}
\def\sol{^{\rm sol}}
\def\tot{^{\rm tot}}

\def\ppp{{\rm path}}
\def\pppp{{\rm paths}}

\def\DS{\Delta S}

\def\inp{^{\rm in}}
\def\out{^{\rm out}}

\section{Introductory remarks}

Over the last about ten to twenty years, stochastic thermodynamics has emerged as a
comprehensive framework for describing small driven systems in contact with 
(or embedded in) a heat bath like colloidal
particles  in laser traps or biomolecules and
biomolecular networks. 
 As an essential concept, the notions of classical
thermodynamics like work, heat and entropy production are identified on the
level of fluctuating trajectories. The distributions of these quantities obey
various universal exact  fluctuation relations.

In the first part of these lecture notes, these concepts will be developed for
a driven system obeying a Markovian dynamics on a discrete set of states which
 implicitly also contains the case of overdamped motion on a continuous state
space  usually described by  Langevin equations. Since this part
is well established by now, only a few selected references to the original
key papers will be given. A more comprehensive guide to the vast literature
concerning refinements and 
theoretical and experimental case studies, can be found, inter alia,
 in several recent reviews 
\cite{jarz11,seif12,vdb15,cili17}.

The second part deals with a more recent development concerning the fluctuations
in non-equilibrium steady states for which a family of inequalities were found
among which the most prominent one constrains the mean and variance of currents
in terms of the overall entropy production. This universal relation can also
be expressed as the inevitable trade-off between cost and precision of any
thermodynamically consistent process which has been
dubbed the thermodynamic uncertainty relation. Its proof follows from a universal
bound on the large deviations of any current. Stronger bounds on these fluctuations
follow with somewhat more knowledge about the driving forces and the topology
of the underlying network. With these relations, measured
fluctuations allow to infer otherwise hidden properties of these systems. 
This presentation is not intended to be  an exhaustive review of these recent
(and ongoing) 
developments but rather a  pedagogical introduction to them.

\section{Closed system in contact with a heat bath}
\label{sec-2}

\subsection{Meso-states}
 \label{sec:meso}
Starting on a very general level, we consider a closed system with micro-states $\{\xi\}$ and 
energy $H(\xi)$ in contact with a heat bath at inverse temperature $\beta$.
In equilibrium, free energy, internal energy and entropy are given by
\beq
F=-(1/\beta)\ln\sum_\xi \exp[{-\beta H(\xi)}], ~~ E=\partial_\beta(\beta F), ~~S=\beta^2\partial_\beta F=\beta(E-F) ,
\label{eq1}
\ee respectively.

We then partition the total phase space into a set of observable meso-states $\{I\}$.
Each micro-state $\xi$ is assumed to belong to one and only one meso-state $I$
to which many micro-states $\xi\in I$ contribute.
  In equilibrium (superscript$^e$), the probability to find the system in meso-state $I$ is then given by
\beq
P_I\eq =\sum_\xai \exp[-\beta(H(\xi)-F)] \equiv \exp[-\beta(F_I-F)]
\label{eq:pie}
\ee
where the last equality defines the free energy $F_I$ of the state $I$. This
identification is justified, first, since the mean energy in state $I$ can be expressed
as
\beq
E_I=\sum_\xai \pxi H(\xi) = \partial_\beta(\beta F_I),
\ee
where  
\beq
\pxi = \exp[{-\beta(H(\xi)-F)}]/P_I\eq=\exp[{-\beta(H(\xi)-F_I)}]
\ee
is the conditional probability for the micro-state $\xi$ given the meso-state $I$.
Second, defining an "intrinsic" entropy $S_I$ from $F_I$ as in ({\ref{eq1})
leads to
\beq
S_I\equiv \beta^2\partial_\beta F_I = \beta(E_I-F_I)=-\sum_\xai\pxi\ln\pxi \equiv S[\pxi],
\label{eq:sint}
\ee
which is the Shannon entropy of the conditional probability.\footnote{Throughout this presentation,
entropy is dimensionless, i.e., Boltzmann's constant is set to 1, and 
$S[p_i]\equiv -\sum_i p_i \ln p_i$ denotes the Shannon entropy of an arbitrary discrete
probability distribution.}
With these expressions, in equilibrium, mean energy, entropy and free energy of the system can also
be written as
\beq
E=\sum_I P_I\eq E_I, ~~~~
S=\left\{\sum_I P_I\eq S_I\right\} + S[P_I\eq],
~~~~ 
F=\left\{\sum_I P_I\eq F_I\right\}-(1/\beta)S[P_I\eq] ,
\ee
respectively.

\subsection{Trajectory, time-scale separation, transition rates and master equation}
In the course of time, 
the system moves along a trajectory $I(t)$ of meso-states. While in
principle any partition into meso-states is formally possible, such a separation 
makes  physical sense, and will lead to stochastic thermodynamics,  if transitions between meso-states are slow while transitions
between the micro-states belonging to one meso-state are fast.
As a necessary
condition, obviously, the heat bath has to relax at least as fast. Ideally, the dynamics along
such a trajectory then becomes Markovian, which means that there is a (constant) transition rate
$\kij$ for the system in state $I$ to jump to state $J$ independent of how long the system
has already been in state $I$ and how it got there.
Under this assumption, the probability to observe the system at time $t$ in state $I$ follows
the master equation
\beq
\partial_tP_I(t)=\sum_J [P_J(t) \kji-P_I(t)\kij]. 
\label{eq:master} 
\ee 
The transition rates $\{\kij\}$ are not arbitrary but have to fulfill certain conditions.
First, under this dynamics, the equilibrium distribution (\ref{eq:pie})
should remain invariant. Second, in equilibrium, there should be no net flow across any
link $(IJ)$ which means that in a long trajectory the number of transitions between $I$ and $J$
should be the same as the number of those between $J$ and $I$. These two conditions imply
that
\beq
\kij/\kji={P_J\eq}/{P_I\eq}=\exp[-\beta(F_J-F_I)]=\exp(-\beta\Dij F)
= \exp(-\beta\Dij E+\Dij S)
\label{eq:locdb}
\ee where we use the notation $\Dij A\equiv A_J-A_I$ throughout for any
function defined on meso-states. This constraint does not fully specify the dynamics.
In order to determine the rates beyond this constraint on their ratio, one would need a
more specific model. It turns out, however, that a number of general results can be derived
that are independent of such non-universal choices.

Crucially, under the assumption of fast equilibration within a meso-state,
the dynamics (\ref{eq:master}) can be used not only in genuine equilibrium but also in situations where
the system has initially been prepared in one meso-state, or, more generally, 
in an initial condition $\{P_I^0\}$  since the future evolution from state $I$ is independent of
whether that state has been prepared initially or has been visited in the course of an equilibrium trajectory.
If the set of meso-states is connected, i.e., does not split into two subsets among which there is
no link, the Perron-Frobenius theorem guarantees that any initial distribution will approach the
unique equilibrium distribution, $P_I(t)\to P_I\eq$ as $t\to \infty$
for all $I$ \cite{vankampen}.
\def\sol{^{\rm sol}}

\subsection{Thermodynamics along a trajectory and in the ensemble}

Along a trajectory $I(t)$, the internal
energy of the system becomes a stochastic quantity, $E(t)=E_{I(t)}$. Since the system is closed,
any energy change of the system is compensated by a corresponding change in the energy of the heat bath, which
can be interpreted as a perpetual exchange of heat along an individual trajectory
as introduced by Sekimoto  \cite{seki98}. Quantitatively, for a transition from $I$ to $J$,
this first law reads
\beq
\Dij E\equiv E_J-E_I = -Q_{IJ} .
\label{eq:firstIJ}
\ee
Here, $Q_{IJ}>0$ corresponds to heat dissipated in the bath thus increasing
its entropy by $\beta Q_{IJ}$. Moreover, the system carries  
entropy 
\beq
S\sys(t) \equiv S_{I(t)}-\ln[P_{I(t)}(t)] .
\label{eq:secondtrajIJ}
\ee
The first part is the intrinsic entropy defined in (\ref{eq:sint}) above whereas the second
is the "stochastic entropy" \cite{seif05a} that can change even while the system
remains in the same meso-state.
Consequently, a transition from $I$ to $J$ at time $t$ entails  the total
entropy change
\beq
\Dij S\tot(t)  = \beta Q_{IJ} +\Dij S\sys(t) = \beta Q_{IJ} +S_J-S_I + \ln[P_I(t)/P_J(t)]= \ln [P_I(t)\kij/P_J(t)\kji] 
\label{eq:stotIJ}
\ee
where we have used (\ref{eq:locdb}) and (\ref{eq:firstIJ}).

This particular identification of a trajectory dependent total
entropy change gains further justification through the
following implications and observations.
First, in equilibrium, the entropy is constant along any trajectory since the various
contributions in (\ref{eq:stotIJ}) add up to zero 
for each transition. This would not be the case if we had not included the term called stochastic entropy.
Second, on the ensemble level, the probability for a transition from $I$ to $J$ at time $t$ is
$P_I(t)\kij$. Consequently, the mean rate of entropy production 
at time $t$ becomes
\beq
\langle \dot S\tot(t)\rangle  \equiv \sum_{IJ} P_I(t)\kij\Dij S\tot(t) = \sum_{I<J}[P_I(t)\kij-P_J(t)\kji]\ln[ P_I(t)\kij/P_J(t)\kji]
\ee as  introduced by Schnakenberg \cite{schn76}. 
Here and throughout, the notation $I<J$ means that each link $(IJ)$ is counted only once. Since 
$(x-y)\ln(x/y)\geq 0$ for all non-negative $(x,y)$, we immediately get the second law
\beq
\langle \dot S\tot(t)\rangle  \geq 0 , 
\label{eq:sec-law}
\ee
independently of the initial condition at any time during the evolution. Note that without the term called above
stochastic entropy, it is easy to conceive a case for which the mean contribution from heat and intrinsic entropy
becomes negative.

Third, as a refinement of the second law (\ref{eq:sec-law}), using the path weight
and the concept of "time-reversal" introduced below in Sect. \ref{sec:pi}, one can easily prove the integral fluctuation theorem 
for total entropy production \cite{seif05a}
\beq
\langle \exp[{-\Delta S\tot}]\rangle = 1 ,
\label{eq:s-int}
\ee where the exponent corresponds to the total entropy change along a trajectory of
arbitrary but fixed length $\t$ and the average is over
the ensemble that evolves from an arbitrary initial condition $\{P_I^0\}$.

Fourth, this identification of entropy along a trajectory can be motivated
from "time-reversal" by refining an argument indicated at in \cite{seif11}. Suppose we postulate the following conditions 
for the  entropy change along a trajectory associated with a transition $I\to J$
at time $t$: (i) The contribution
$\Delta S\tot_{IJ}$ from this jump is the negative of a putative contribution of the reversed
jump taking place at the same time, $\Delta S\tot_{IJ}(t)=-\Delta S\tot_{JI}(t)$. (ii) The mean
rate of total entropy production is non-negative at any time $t$ for any initial distribution  $\{P_I^0\}$.
It is then straightforward to show that 
$\Dij S\tot(t) $ should be of the form $g(\ln[ P_I(t)\kij//P_J(t)\kji])$ with
$g(y)=-g(-y)$. If one imposes additionally that along a trajectory
the total entropy production is additive in system and bath, with the latter given by $\beta Q_{IJ}$, $g(y)$ must be linear
leading to (\ref{eq:stotIJ}) up to a multiplicative constant. 

\subsection{Time-dependently driven system}
\label{sect:driven}
\def\it{_{|I(t)}}
The framework discussed above can easily be adapted to the situation where one assumes
that the system is driven externally leading to a time-dependence of the microscopic
Hamiltonian as $H(\xi,\l)$ where $\l(t)$ is a time-dependent control parameter
\cite{jarz97,jarz97a,croo99}. This
time-dependence should be slow enough so that the micro-states within each meso-state 
can still equilibrate. As a
first consequence, free energy, internal energy and entropy of the meso-states
as defined in (\ref{eq1}) above (here with $H(\xi)\to H(\xi,\l)$) become
time-dependent, $F_I(\l), E_I(\l), S_I(\l)$. Second, thermodynamic consistency
now requires that the ratio of the transition rates (\ref{eq:locdb}) becomes time-dependent and is given by
\beq
{\kij(\l)}/{\kji(\l)}=\exp[-\beta\Dij F(\l)] 
\ee with $\l=\l(t)$.

In such a setting, along the trajectory $I(t)$ of meso-states the rate of
work applied to the system
should be identified as
\beq
\dot W(t)=\sum_{\xi\in I(t)}\exp[-\beta(H(\xi,\l)-F_I(\l))](\partial H/\partial \l)\dot \lambda=\partial_\l F_I(\l)\it \dot \l,
\label{eq:wdot}
\ee which corresponds to the appropriately averaged change of the microscopic
Hamiltonian. Here, and throughout along
trajectories, the dot denotes a total time-derivative (possibly delta-like due to jumps). Note that this expression differs slightly from earlier identifications
of work, $ \partial_\l E_I(\l)\it \dot \l$, for a stochastic dynamics \cite{jarz97a,croo99} since we allow the meso-states to have different intrinsic entropy \cite{seif11,seki07}. 

The first law then allows us to identify the rate of dissipated
heat as
\beq
\dot Q(t)\equiv \dot W(t)-\dot E(t) =  \dot W(t) -\sum_J \dot \delta_{JI(t)}E_J(\l)
-\partial_\l E_I(\l)\it \dot \l
=-\sum_J \dot \delta_{JI(t)}F_J(\l)-\frac {d}{dt} S_{I(t)}(\l)/\beta  .
\label{eq:qdot}
\ee
Since internal energy and  intrinsic entropy of a meso-state can become time-dependent, these expressions show
that, in contrast to the case without driving, heat is now exchanged even if
the system stays in the same meso-state.

The total entropy change along a trajectory becomes
\beq
\dot S\tot(t)  = \beta \dot Q(t)+ \frac{ d}{dt}[S_{I(t)}(\l)-\ln P_{I(t)}(\l)]. 
\ee
The integral fluctuation theorem for
entropy production ({\ref{eq:s-int}) holds for arbitrary driving $\l(t)$ \cite{seif05a} as does, consequently, the
second law on the ensemble level. 

\section{From a closed to an open system}
\subsection{Transition rates}

So far, we have considered a closed system in contact with a heat bath. For systems involving
transport (possibly against an external force) and/or chemical reactions, it is advantantageous
to split this system further into (i) a core system of interest, 
(ii) the surrounding solution, which will effectively act as a particle reservoir,
and (iii) a part responsible for providing an external mechanical force. The latter two parts will be associated
with external driving and, possibly, with the extraction of chemical or mechanical work.

\begin{figure}[htb]
	\centering
 	\includegraphics[scale=.4]{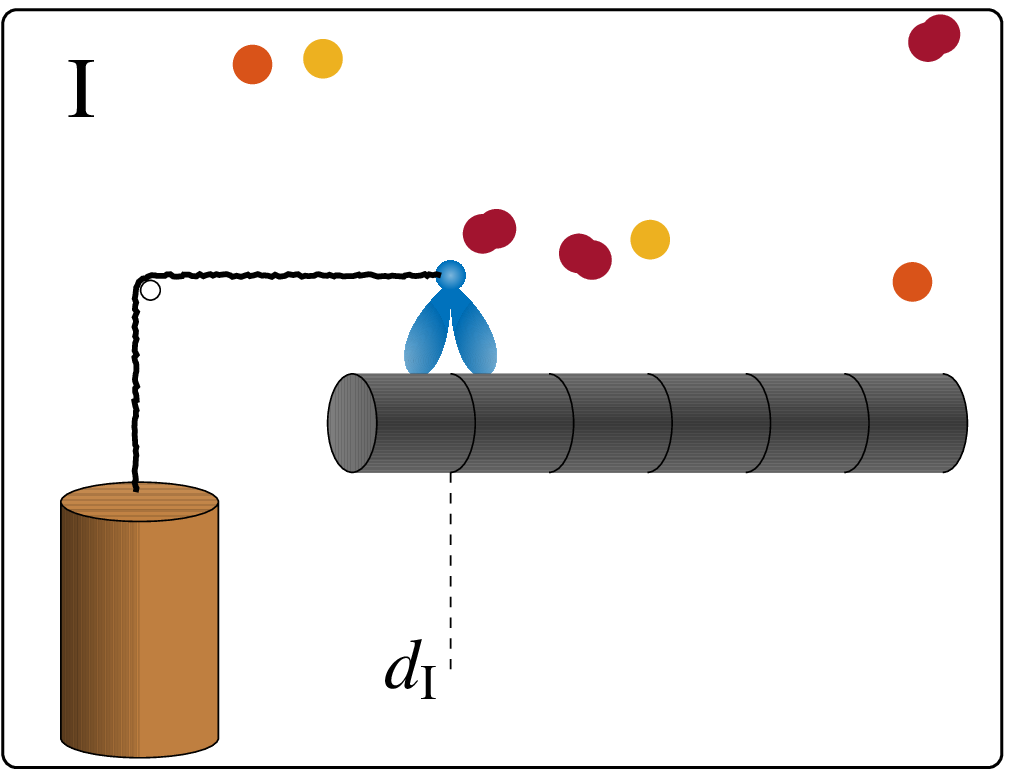}~~~
 	\includegraphics[scale=.4]{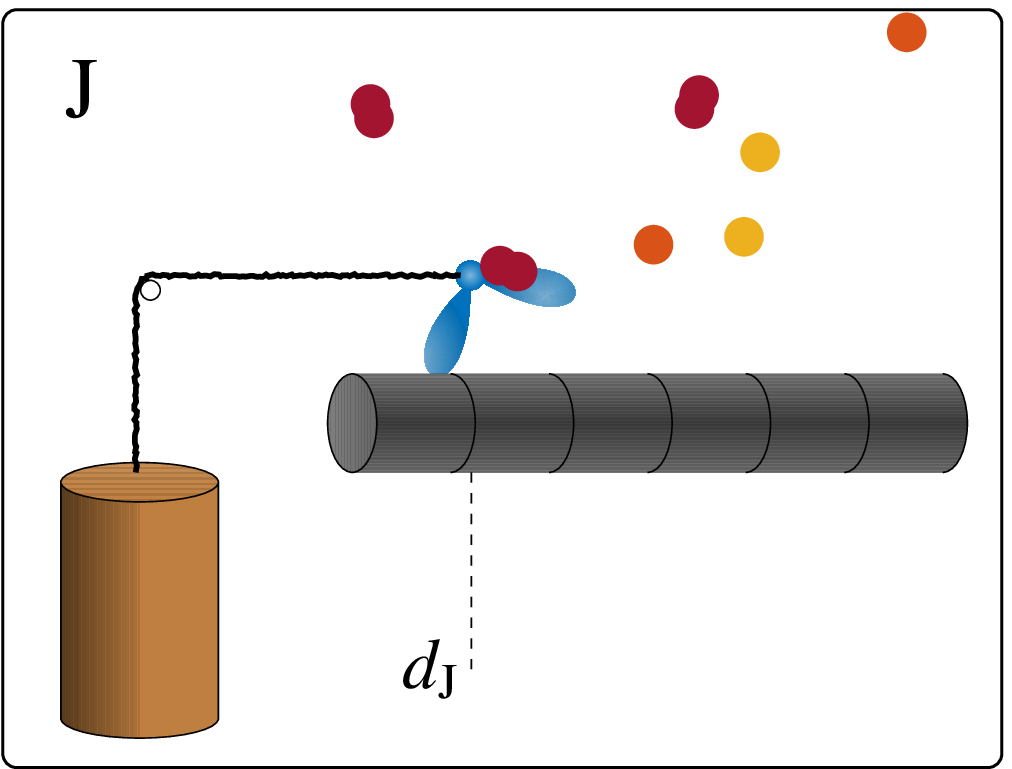}~~~
 		\includegraphics[scale=.4]{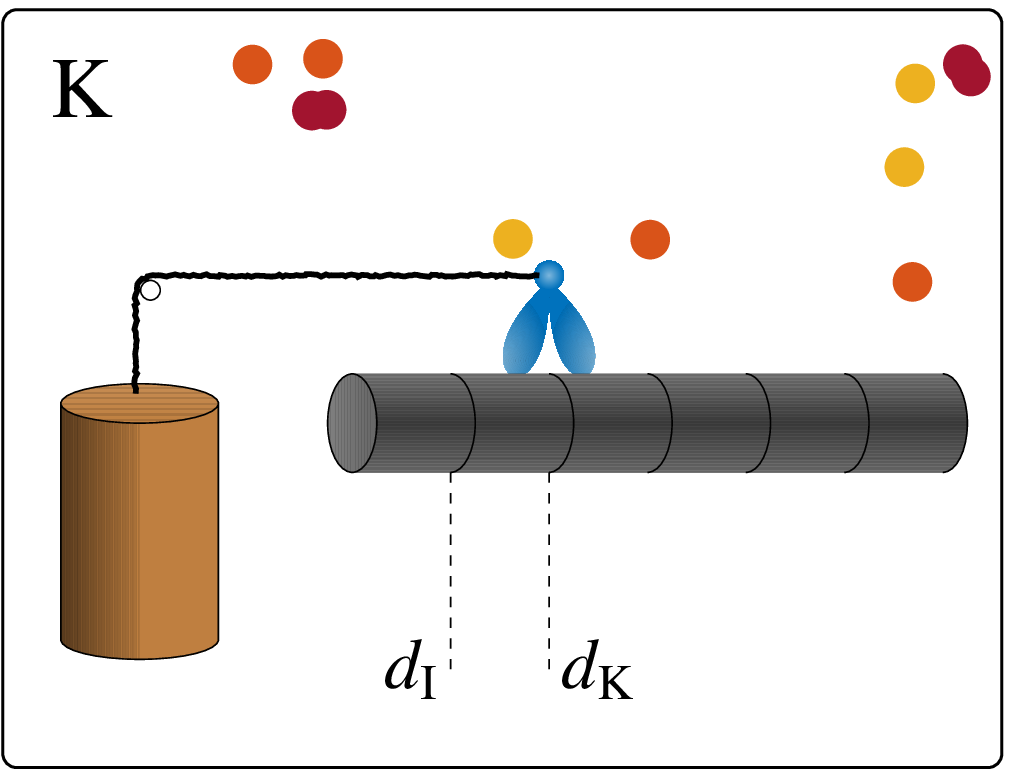}

	\caption{Scheme of a molecular motor 
	 powered by the hydrolysis of an ATP
	shown here as ATP (red) $\to$ ADP (orange)  + P (yellow) stepping along a filament against
	a force represented by a weight. From the closed system's perspective,
	the mesostates $I,J,K$ contain the state of the enzyme $i_{I,J,K}$, the
	total number of molecules of each species, and the position of the motor relative
	to the fixed track, $d_{I,J,K}$. In the open system's perspective of
	just the motor, one can focus on the states of the motor, here
	oversimplified as just two different ones
	$i_I=i_K$ (tightly bound to the track) and $i_J$ (with one "head" loose 
	binding to ATP).}
	\label{fig:mm}
\end{figure}

As the core  system, we consider paradigmatically an enzyme or a molecular motor 
(or several of them) that induce enzymatic reactions, like the hydrolysis of ATP, between solutes of
various species $\alpha$ in the solution, see Fig. \ref{fig:mm}. Consequently,  to each meso-state $I$
of the full system there corresponds a state of the enzyme(s) $i_I$ including tightly bound solutes. 
The mapping from $I$ to $i_I$ is unique with many meso-states
$I$ leading to the same $i_I$.

The change in free energy difference of the full system upon a transition from $I$ to $J$ can then be written as
\beq
\Dij F = F_{i_J}-F_{i_I}-\sum_\alpha\mu^\alpha\Delta_{{I}{J}}N^\alpha +fd_{{I}{J}}.
\ee
The first term is the free energy difference of the two enzyme configurations. If these
two configurations contain a different number of bound solutes, 
$\Delta_{{I}{J}}N^\alpha\not = 0$, the second term quantifies the
free energy difference of the surrounding solution which we have characterized by a set
of chemical potentials $\{\mu^\alpha\}$ that are essentially determined by the
respective concentrations of the various species.\footnote{This identification can be made more formal
as discussed in \cite{seif11}.} Likewise, if the transition $I\to J$ involves 
 the motor stepping  a distance $d_{{I}{J}}\equiv d_J-d_I$ against the external force $f$, the
last term is the corresponding free energy change. 
We assume that there are no transitions that change the numbers of free solutes without
a concomitant change of the enzyme configuration. This means that there are no chemical
reactions taking place in the solvent that are not enzymatically induced.
We can then replace all
reference to the original meso-states $IJ$ by considering the chemical
potentials and the force to be given and write $IJ\to ij $.
The transitions between the internal states of the motor or enzyme (including binding and 
release of solutes) must then obey the constraints
\beq
{k_{ij}}/{k_{ji}}=\exp[-\beta(\dij F -\sum_\alpha\mu^\alpha \dij N^\alpha + f d_{ij})] .
\label{eq:det-bal-ness}
\ee

The core system has thus become  an open system connected to a heat bath with inverse
temperature $\beta$ and chemostats with chemical potential $\{\mu^\alpha\}$
possibly subject to an external force $f$.
The master equation (\ref{eq:master}) expressed for the probability of just the
core states reads
\beq
\partial_t p_i(t) = \sum_j[-k_{ij}p_i(t) + k_{ji}p_j(t)].
\label{eq:master-ness}
\ee

\subsection{Thermodynamics along trajectories and in the ensemble}

For this open system, 
the first law (\ref{eq:firstIJ}) along a transition $i\to j$  becomes
\beq
 \dij E + \Delta_{ij} E\sol +  W_{ij}^{\rm out} = -Q_{ij} .
\label{eq:first-trajij}
\ee
On the left hand side, the first term is the energy change of the enzyme, the second one the energy
change in the surrounding solution, formally the reservoirs, if the two core states
contain a different number of bound solutes, and the third term the extracted
mechanical work, $W_{ij}^{\rm out}\equiv fd_{ij}$ delived against an external force $f$ in a transition $i\to j$.
 These three different contributions to what would be the internal
energy in a description as a closed system must be compensated by the dissipated heat since the
total energy, including that of the heat reservoir, must be conserved.

The entropy change (\ref{eq:secondtrajIJ}) of the system, now consisting of core system and solution, that is 
 associated with a transition $i\to j$,
 becomes
\beq
\dij S\sys(t)\equiv \dij S +\Delta_{ij} S\sol + \ln[p_i(t)/p_j(t)].
\label{eq:secondtrajij}
\ee
The first term contains the change in intrinsic entropy of the enzyme, the second one the
entropy change of the surrounding solution, the last one the stochastic entropy of the
enzyme. Note that there is no more stochastic entropy associated with the state of the
solution since the reservoir is fully characterized by the chemical potentials. Likewise,
there is neither intrinsic nor stochastic entropy associated with a putative mechanical work source.
With (\ref{eq:det-bal-ness}), the total entropy change can be written as
\beq
\dij S\tot(t)= \beta Q_{ij}+ \dij S\sys(t)=\beta Q_{ij}+ \dij S +\Delta_{ij} S\sol + \ln[p_i(t)/p_j(t)]= \ln[p_i(t)k_{ij}/p_j(t)k_{ji}]  .
\label{eq:secthree}
\ee

On the ensemble level, the first and second law now become
\beq
\langle \dot E(t)\rangle +\langle \dot E\sol(t)\rangle  + \langle\dot W\out(t)\rangle = - \langle \dot Q(t)\rangle
\ee and 
\beq
\langle \dot S\tot(t)\rangle  = \sum_{ij} p_i(t)k_{ij}\dij S\tot(t) = 
\sum_{i<j}[p_i(t)k_{ij}-p_j(t)k_{ji}]\ln[p_i(t)k_{ij}/p_j(t)k_{ji}] \geq 0,
\label{eq:secens}
\ee
respectively.
The integral fluctuation theorem (\ref{eq:s-int}) for total entropy production holds unmodified.

\subsection{Non-equilibrium steady states (NESSs)}

The master equation (\ref{eq:master-ness}) 
with the thermodynamic consistency condition (\ref{eq:det-bal-ness}) will typically approach a unique stationary state,
$\{p_i(t)\}\to \{p\ss_i\}$ as $t\to\infty$ independent of the initial distribution $\{p_i^0\}$ \cite{vankampen}.
This stationary distribution can either be calculated as the right eigenvector to the eigenvalue
0 of the corresponding matrix or obtained from a nice graphical construction explained, e.g., in \cite{zia07}, which works particularly
well for small networks.

In this non-equilibrium steady state, there will  be net currents across
some links,
\beq
j^s_{ij}=p^s_ik_{ij}-p^s_jk_{ji}\not = 0,
\ee
which distinguishes such a NESS fundamentally from genuine equilibrium. In a
NESS, the mean rate of entropy production (\ref{eq:secens}) (denoted by $\sigma$ from now on) is constant and given by
\beq
\sigma= 
\sum_{i<j}[p_i^sk_{ij}-p_j^sk_{ji}]\ln[p_i^sk_{ij}/p_j^sk_{ji}] \geq 0.
\label{eq:sigma}
\ee

\subsection{Remark on strong coupling and fixed pressure}
So far, we have implicitly assumed that the coupling between the system and the heat bath is
weak.  As shown in \cite{seif16}, a thermodynamically consistent identification of trajectory dependent internal energy and entropy
is, however, possible even without this assumption, which, for biomolecular systems, will not
necessarily hold. Likewise, for biochemical systems, the assumption that
the system (including the particle reservoirs) has a fixed
volume should typically be replaced by 
assuming a fixed pressure $\cal P$. However, all definitions
and identifications on the trajectory level from the above sections remain valid, provided free energies differences $F_j-F_i$ are
replaced by Gibbs free energy differences $G_j-G_i=F_j-F_i + {\cal P} (V_j-V_i)  $. 
The exception is the first law (\ref{eq:first-trajij}), which now reads
\beq
  E_j-E_i + \Delta_{ij} E\sol + {\cal P}(V_j-V_i)+ W\out_{ij} = -Q_{ij} ,
\ee
 with  the concomitant identification
of heat. In order not to overburden the presentation, we stick to the weak coupling and
constant volume
case in the following and refer to \cite{jarz17} for an instructive discussion of
the latter and related aspects.

\section{"Time-reversal", entropy production, and fluctuation relations}
\label{sec:pi}
For this section, we first return to a general open system characterized by a set of states
$\{i\}$ with time-independent transition rates (\ref{eq:det-bal-ness}).

\subsection{Path weight for a trajectory}
The probability $p[i(t)|i_0]$ to observe a trajectory $i(t)$ starting at time $t=0$ in $i(0)=i_0$
and jumping at times $t_j$ from $i_j^-$ to $i_j^+$ ending up 
after $J$ jumps at time $t=\t$ in $i(t)=i_\t$ is given by\footnote{This 
expression arises from applying repeatedly a straightforward generalization
of the fact that if an event occurs with a rate $k$, the probability that
it occurs for the first time at time $t$ is $p(t)=\exp(-kt)k$ given by a
product of a waiting probability ("nothing happens") and the rate.
}
\beq
p[i(t)|i_0] =\left\{ \prod_{j=1}^J \exp[-r_{i_j^-}(t_j-t_{j-1})]k_{i_j^-i_j^+}\right\}
\exp[-r_{i_\t}(\t-t_J)] =\exp\left[-\sum_i r_i\tau_i\right] \prod_{ij} k_{ij}^{n_{ij}}
\label{eq:pathi}
\ee
where the last product runs over all links (in both directions). Here, 
\beq
r_i\equiv \sum_jk_{ij}
\ee
is the escape rate of state $i$. For $J=0$, i.e., the trajectory without any jump, the term
 in curly brackets should be set to 1 and, in the remainder,  $t_J$ to 0  leading to the weight $\exp[-r_{i_0}\t]$
 for this trajectory.
The second equality shows that the weight of any trajectory that starts at $i_0$ is fully determined by knowing
the total time $\tau_i$ it spends in a state $i$ and  the number of transitions $n_{ij}$ from $i$ to $j$. Of course, there are many different trajectories leading to the same set  
$\{\tau_i\},\{n_{ij}\},$ which are, in general, not easily summed (or integrated) up.

\subsection{Time-reversed trajectory and time-reversed process}

An important concept for deriving fluctuation relations is 
the time-reversed or "backward" trajectory, or path, 
 $\widetilde \ppp \equiv \tilde i(t)\equiv i(\t-t)$,
running from $\tilde i_0=i_\t$ to $\tilde i(\t)=i_0$. The ratio between the probability
to observe this trajectory given its initial value  and the original ("forward") one follows from (\ref{eq:secthree}) and (\ref{eq:pathi}) as
\beq
\frac{p[\widetilde\ppp|\tilde i_0]}{p[\ppp|i_0]}\equiv
\frac {p[\tilde i(t)|\tilde i_0]}{p[i(t)|i_0]}
=\exp[-\sum_{ij}n_{ij} \ln(k_{ij}/k_{ji})]=\exp[-(\beta Q[\ppp]+ \Delta S[\ppp])],
\label{eq:s-ratio-0}
\ee
since in the path weight the terms involving the escape rates are identical for both paths. This ratio is hence given by the heat dissipated along the original path and
 the concomitant change in intrinsic entropy.

In a useful generalization, while always drawing the forward path from the original 
distribution $\{p_i^0\}$, 
the backward one can be drawn from a, in general, fictitious ensemble $\{\tilde p_{\i}^0\}$ not necessarily 
 given by $\{p_i(\t)\}$, where the latter would be the final distribution along the forward process.
Denoting this probability distribution for the backward paths by $\tilde p[\widetilde \ppp]$,
we get easily
\beq
\frac{\tilde p[\widetilde\ppp]}
{p[\ppp]}
=\frac{\tilde p[\widetilde\ppp|\tilde i_0=i_\t]\tilde p_{i_\t}^0}
{p[\ppp|i_0]p^0_{i_0}}
=\exp[-(\ln[p_{i_0}^0/\tilde p_{i_\t}^0]+\beta Q[\ppp]+ \Delta S[\ppp])] .
\label{eq:s-ratio-1}
\ee
This master relation is a useful starting point for a unified
derivation of several famous non-equilibrium relations as discussed in the following.

\subsection{Fluctuation theorem for entropy production in a non-equilibrium steady state (NESS)}
For a NESS, we can draw the initial state for forward and backward path from the stationary distribution $\{p_i^s\}$. Since the first term in the exponent of
(\ref{eq:s-ratio-1}) then becomes the change in stochastic entropy along the path,
we get 
\beq
\frac{p[\widetilde\ppp]}{p[\ppp]}
=\exp[-\DS\tot[\ppp]] .
\label{eq:s-ratio-3}
\ee
Hence, in a NESS the probability to observe the time-reversed trajectory
compared to the original one is exponentially small in the total entropy production along the original path.
For a NESS, this relation quantifies an often somewhat imprecisely insinuated relation
between  entropy production and 
the "breaking" of time-reversal symmetry.

This behavior under time-reversal implies a remarkable symmetry of the distribution
$p(\DS\tot)$  of total entropy production in a NESS, called the fluctuation theorem (FT),  since
\begin{eqnarray}
p(-\DS\tot)&=&\sum_\pppp p[\ppp]\delta(\DS\tot[\ppp]+\DS\tot)\nn\\
&=&\sum_\pppp p[\widetilde \ppp]\exp[\DS\tot[\ppp]]\delta(\DS\tot[\ppp]+\DS\tot)\nn\\
&=&\sum_{\widetilde \pppp}p[\widetilde \ppp]
\exp[-\DS\tot[\widetilde \ppp]]\delta(-\DS\tot[\widetilde\ppp]+\DS\tot)\nn\\
&=&\exp[-\DS\tot]p(\DS\tot) .
\end{eqnarray}
Here, the first equality is the definition of the probability distribution, in the second we use the
ratio (\ref{eq:s-ratio-3}), in the third the anti-symmetry $\DS\tot[\ppp]=-
\DS\tot[\widetilde\ppp]$ 
and the fact that summing over the reversed paths is exhaustive. Due to the
inclusion of stochastic entropy, this
relation holds as shown here even for a finite total time $\t$ \cite{seif05a}.
Without this term, it has been first derived for stochastic dynamics in the
long-time limit in \cite{kurc98,lebo99}. Earlier versions have been derived for thermostatted and chaotic dynamics \cite{evan93,evan94,gall95}. 

\subsection{Time-dependent driven systems: Jarzynski and Crooks relation}

For a closed system connected to a heat bath and driven by a time-dependent 
Hamiltonian $H(\xi,\l)$   as introduced in
Sect. 
\ref{sect:driven} above, the weight for a trajectory is slightly more involved
than the "simple" expression (\ref{eq:pathi}). First, since the escape rate becomes time-dependent,
the respective terms in the exponent must be replaced by time-integrals. Second,
the weight now depends on the times when the transitions have taken place rather
than just on their numbers. Moreover, the reversed process now also involves
time-reversal of the control parameter according to $\tilde \l(t)\equiv \l(\t-t)$.

It is a simple exercise to show that the (inverse) ratio of probabilities of observing
the original trajectory under the forward driving and the probability of observing
the time-reversed one under the time-reversed driving starting with arbitrary initial
conditions is still given by (\ref{eq:s-ratio-1}) since the crucial time-dependences cancel. For this closed driven system, we should use 
(\ref{eq:s-ratio-1}) with capital letters for meso-states
and probabilities which was the notation in Sect. \ref{sec-2}. 

For a system that is driven from an initial parameter $\l_0$ to a final  $\l_1$,
by starting the original process in thermal equilibrium and the backward one
also in the respective thermal equilibrium, $P^e_I(\l)=\exp[-\beta(F_I(\l)-F(\l))]$, the first term in the exponent
of (\ref{eq:s-ratio-1}) becomes
\beq
\ln[p_{i_0}^0/\tilde p_{i_\t}^0]\mapsto
\ln[P_{I_0}^e(\l_0)/ P_{I_\t}^e(\l_\t)]=\beta [-F_{I_0}(\l_0)+F(\l_0)+F_{I_\t}(\l_\t)-F(\l_\t)]=\beta[\Delta F[\ppp]-\Delta F],
\ee where $F(\l)$ denotes the free energy at control parameter $\l$
and $\Delta F\equiv F(\l_\t)- F(\l_0)$
is the free energy difference of the
system at the two values of the control parameter. After integrating (\ref{eq:wdot}) and (\ref{eq:qdot}) 
along a trajectory, the sum of second and third term  in the exponent
of (\ref{eq:s-ratio-1}) becomes  
\beq
\beta Q[\ppp]+ \Delta S[\ppp] =\beta (W[\ppp]-\Delta F[\ppp]) .
\ee
Putting everything together, one gets
\beq
\tilde p[\widetilde \ppp]/p[\ppp]=\exp[-\beta (W[\ppp] -\Delta F)] .
\ee
  By repeating 
essentially the same
calculation as above for the derivation of the FT in a
NESS, one  gets the Crooks relation \cite{croo99}
\beq
\tilde p(-W)=p(W) \exp[-\beta (W -\Delta F)] .
\label{eq:crooks}
\ee
Consequently, the free energy difference of two states can be determined by
measuring the crossing point of the work distributions using the original and
the time-reversed protocol. For the beautiful first experimental application of this relation
with biomolecules, see \cite{coll05}.

Finally, and here certainly not following the original history, one gets the
famous Jarzynski relation \cite{jarz97,jarz97a} by integrating (\ref{eq:crooks}) over $W$ as
\beq
\langle \exp[-\beta W]\rangle = \exp[-\beta \Delta F] ,
\ee
whose manifold consequences are authoratively reviewed in \cite{jarz11}.
 
Note that we have derived the Jarzynski and the Crooks relation here using
stochastic dynamics and meso-states that possess intrinsic entropy. The
original derivation of the former \cite{jarz97} used Hamiltonian dynamics for the closed system and
coupling and decoupling from a heat bath. Likewise, the original (and many present)
derivations using stochastic dynamics \cite{jarz97a,croo99}
ignore the intermediate consequences 
of intrinsic entropy, which are no longer explicitly visible  at the end anyway.

\section{Asymmetric random walk as a simple thermodynamically consistent paradigm}
\subsection{Model}
\label{sect:arw}
For a simple asymmetric random walk, we introduce a few concepts that will be explored for more general systems in the following sections. This model can also
serve as a simple
description of a molecular motor running along a filament. 
In each step of length $d$, the motor works against  an external force $f$
and is powered by hydrolysis of one molecule of ATP 
that provides $\Delta \mu$ of
free energy in each forward reaction (ATP $\to$ ADP + P$_i$) and generates the same amount in a backward reaction. Thermodynamic consistency (\ref{eq:det-bal-ness}) demands for the ratio
of forward, $k_+$, to backward, $k_-$, rate 
\beq
k_+/k_-=\exp[\beta(\Delta \mu -fd)]\equiv \exp A,
\ee
which defines the (dimensionless) affinity $A$.

\subsection{Fluctuations}
After a time $t$, the motor has made $n_+$ steps in the forward and
$n_-$ steps in the backward direction. Their probability distribution
obeys
\beq
\partial_tp(n_+,n_-,t)=-(k_++k_-)p(n_+,n_-,t)+k_+p(n_+-1,n_-,t)+k_-p(n_+,n_--1,t).
\ee
Since the steps in the two directions correspond to two independent Poisson processes,
this distribution function is simply
\beq
p(n_+,n_-,t) = [(k_+t)^{n_+}/n_+!] [(k_-t)^{n_-}/n_-!]\exp[-(k_++k_-)t],
\label{eq:p-arw}
\ee
as is easily verified a posteriori by insertion.\footnote {Note that for this 
simple asymmetric random walk the summation of (\ref{eq:pathi}) over all
transition times is obviously possible leading to the additional
factor $t^{n_+n_-}/(n_+!n_-!)$ when (\ref{eq:p-arw})  is derived from (\ref{eq:pathi}).}
Mean value and dispersion are given by
\beq
\langle n_\pm\rangle = k_\pm t {\rm ~~and~~} \langle(n_\pm-\langle n_\pm\rangle)^2\rangle = k_\pm t .
\ee
For the net number of steps in forward direction, $n\equiv n_+-n_-$, one gets
for mean and dispersion
\beq
\langle n\rangle = (k_+-k_-)t {\rm ~~and~~} \langle(n-\langle n\rangle)^2\rangle =(k_++k_-)t\equiv 2Dt 
\ee
with the diffusion constant $D\equiv (k_++k_-)/2$.

The mean rate of entropy production (\ref{eq:sigma})
becomes
\beq
\sigma  =(k_+-k_-)\ln(k_+/k_-)=j^sA 
\label{eq:s-arw}
\ee
with the mean net current $
j^s\equiv \nu_+^s-\nu_-^s=k_+-k_- .$

For large $t$, the factorials in (\ref{eq:p-arw}) 
can be approximated by Stirling's formula leading to
\beq
p(n_+,n_-,t)\approx \exp\{-t[\nu_+\ln(\nu_+/k_+)-\nu_+ + k_+ + \nu_-\ln(\nu_-/k_-)-\nu_-+k_-] \} 
\equiv \exp[-t I(\nu_+,\nu_-)]
\label{eq:arw-rate}
\ee
where we have defined the fluctuating rates of directed transition
\beq
\nu_\pm\equiv n_\pm/t 
\ee
with stationary mean value $\nu_\pm^s=k_\pm$ and identified a "rate" function $I(\nu_+,\nu_-)$.

\subsection{Thermodynamic uncertainty relation}

The expressions just given allow for another interpretation in terms of the
precision of a biomolecular process. 
After a time $t$, the motor has "produced"  a number of steps $n$. The uncertainty of
this process is defined as
\beq
\eps^2\equiv \langle (n-\langle n\rangle)^2\rangle/\langle n\rangle^2 = 2D/{j\ss}^2t ,
\label{eq:eps-arw}
\ee
which is a measure of its precision.
During this time $t$, on average, running this process has  generated
$C\equiv \sigma t$ entropy,  which is the (dimensionless) free energy that is not recovered as mechanical work, i.e., the net
thermodynamic cost of the process.
By combining (\ref{eq:s-arw}) and (\ref{eq:eps-arw}) one gets
\beq
C\eps^2=2\sigma D/{j\ss}^2=A \coth (A/2) \geq 2.
\ee 
The product of loss and precision is thus given 
by a function of the affinity. Independently of the value of this affinity,
this product is bounded by, 2 which has been dubbed the "thermodynamic uncertainty
relation" \cite{bara15}. The longer the motor runs the higher the precision, which is a consequence of
the diffusive behavior. On the other hand, the cost increases linearly in time
which implies that the product $C\eps^2$ is time-independent. 
A  higher precision inevitably comes at a higher cost. The inequality is saturated
for vanishing affinity, i.e., close to equilibrium, and 
also close to the stall force, $f\simeq \Delta \mu/d$.
The a priori surprising fact is that the thermodynamic uncertainty relation holds
in a much more general formulation for any thermodynamically consistent Markov process as we  will see below.

\section{Fluctuations in a non-equilibrium steady state (NESS) in the long-time limit}

From now on, we focus on the fluctuations in a NESS in the long-time limit
for which general results can be derived based on techniques from large-deviation theory
as reviewed in \cite{touc09,laza15,bara15d} and by Touchette in this volume.

\subsection{Empirical density, current and traffic}
For a Markov process on an arbitrary set of states, we define two classes
of observables whose mean is extensive in time.
First, there is the (residence or sojourn) time $\tau_i$ a trajectory spends
in the state $i$ with the mean $\langle \tau_i\rangle = p\ss_it $.
It will
become 
convenient to consider the empirical density
\beq
p_i\equiv \tau_i/t,
\ee
whose mean is the stationary distribution $\langle p_i\rangle = p_i^s$

Second, from the number of jumps $n_{ij}$, we get the fluctuating, or empirical, currents and
traffic, defined as 
\beq
j_{ij}\equiv (n_{ij}-n_{ji})/t
{\rm ~~~ and~~~}
t_{ij}\equiv (n_{ij}+n_{ji})/t,
\ee
with mean values
\beq
 j\ss_{ij}=p\ss_ik_{ij}-p\ss_jk_{ji}
 {\rm ~~~ and~~~}
t\ss_{ij}=p\ss_ik_{ij}+p\ss_jk_{ji},
\ee
respectively.
For an arbitrary
current 
\beq j_\alpha\equiv \sum_{ij} n_{ij}d^\alpha_{ij}/t ~~~{\rm with~~ mean~~~}j_\alpha\ss=\sum_{ij} p\ss_i k_{ij}d^\alpha_{ij} 
\label{eq:curr}
\ee the generalized distances $
d_{ij}^\alpha=-d_{ji}^\alpha$
determine how much each transition $i\to j$ contributes. 
A prominent current is the one of total entropy production $j_\sigma $ 
for which $
d^\sigma_{ij}\equiv \ln[p\ss_i k_{ij}/p\ss_j k_{ji}], 
$ whose mean is given by the entropy production rate $\sigma$
(\ref{eq:sigma}).

\subsection{Level 2.5 rate function and contractions}
There is an elegant approach to determine the probability
of large fluctuations, i.e., large deviations from the average
behavior, in the long-time limit. Let $p(\{\tau_i\},\{n_{ij}\},t)$ be the probability
(density) of observing the  residence times $\{\tau_i\}$
and $\{n_{ij}\}$ transitions from $i$ to $j$,  after a time $t$. This probability can be
calculated by introducing an auxiliary set of rates on the
same network of states for which these values would correspond
to the mean behavior. Specifically, for the rates $\hat k_{ij}\equiv n_{ij}/\tau_i$
the stationary distribution becomes
$\hat p_i=\tau_i/t$ and the mean number of transitions is
$\hat n_{ij} =\hat p_i \hat k_{ij}t =n_{ij}$. We denote the
escape rates for these modified rates as $\hat r_i$.

Using this auxiliary set of rates, the ratio of probabilities
of observing the fluctuation $\{\tau_i\},\{n_{ij}\}$
in the original network and in the one with the auxiliary set of rates
 follows from the path integral expression (\ref{eq:pathi}) as \cite{maes08}
\begin{eqnarray}
\frac{p(\{\tau_i\},\{n_{ij}\},t|\{k_{ij}\},i_0)}
{p\{\tau_i\},\{n_{ij}\},t|\{\hat k_{ij}\},i_0)}&=& 
\exp\left[-\sum_i\tau_i(r_i-\hat r_i)+\sum_{ij}n_{ij}\ln(k_{ij}/\hat k_{ij})\right],
\end{eqnarray}
where we introduce a finally irrelevant common initial state $i_0$.\footnote{
 Here and in the following, the advantage of including the denominators on the left hand sides is that one avoids introducing the
Radon-Nikodym derivative for measures and still deals with dimensionless quantities
when later taking logarithms.}
By multiplying with the denominator, summing over initial states and normalizing with the typical fluctuation of the original network, 
 the ratio of probabilities
of observing the
(large) fluctuation and a typical one for the original set
of rates now follows as 
\begin{eqnarray}
\frac{p(\{\tau_i\},\{n_{ij}\},t|\{k_{ij}\})}
{p(\{p\ss_it\},\{p\ss_ik_{ij}t\},t|\{k_{ij}\})}&=& 
\exp\left[-\sum_i\tau_i(r_i-\hat r_i)+\sum_{ij}n_{ij}\ln(k_{ij}/\hat k_{ij})\right]\times
\frac{\sum_{i_0}p(\{\hat p_i t\},\{\hat p_i\hat k_{ij}t\},t|\{\hat k_{ij}\},i_0)p\ss_{i_0}}
{\sum_{i_0}p(\{p\ss_it\},\{p\ss_ik_{ij}t\},t|\{k_{ij}\},i_0)p\ss_{i_0}}.
\end{eqnarray}
The last factor involves the ratio of probabilities of observing  the typical behavior
for the respective set of rates weighted with the probability of the initial state in
the original set of rates. For large $t$, the latter dependence vanishes and the ratio becomes a time-independent function of both sets of rates in leading order.
Consequently, in the long-time limit, the logarithmic ratio of these probabilities
can be written in the form 
\beq
-\lim_{t\to \infty}(1/t)\ln\left(\frac{p(\{\tau_i\},\{n_{ij}\},t|\{k_{ij}\})}
{p(\{p\ss_it\},\{p\ss_ik_{ij}t\},t|\{k_{ij}\})}\right)= I(\{\tau_i/t\},\{n_{ij}/t\})
\ee
with the rate function
\beq
I(\{\tau_i/t\},\{n_{ij}/t\})=\sum_i(\tau_i/t)(r_i-\hat r_i)+\sum_{ij}(n_{ij}/t)\ln(\hat k_{ij}/k_{ij}) .
\label{eq:i5}
\ee

The auxiliary rates have now served their purpose and we can focus on fluctuating quantities for the
original set of rates. Specifically, we consider the empirical densities, currents and traffics.
Expressed in these quantities, the rate function (\ref{eq:i5}) reads
\beq
I(\{p_i\},\{j_{ij}\},\{t_{ij}\})=\sum_{ij}\left\{ \frac{j_{ij}+t_{ij}}{2}
\left[\ln\frac{j_{ij}+t_{ij}}{2p_ik_{ij}}-1\right]+p_ik_{ij}\right\} .
\ee
This rate function is a very general one, called "level 2.5", which depends on  all empirical densities,  currents and traffics. It is crucial to note that probability
conservation provides a set of constraints
$
\sum_j j_{ij}=0
$
 for all states $\{i\}$, which will be assumed implicitly in the following.


Often one is interested in the corresponding rate function
of a subset of these quantities, or a set of certain functions of them. Such a rate function
can be obtained from the full one through "contraction". Specificially, the rate function for
$f=f(\{p_i\},\{j_{ij}\},\{t_{ij}\})$ is obtained through
\beq
I(f) = {\min_{\{p_i\},\{j_{ij}\},\{t_{ij}\}}}_{|{f(\{p_i\},\{j_{ij}\},\{t_{ij}\})=f }} 
I(\{p_i\},\{j_{ij}\},\{t_{ij}\}) 
\ee since for $t\to\infty$ one can focus on the most likely fluctuation for given constraints.
In general, this constrained minimization cannot be performed analytically. An upper bound on the rate
function for $f$, however, can be obtained by inserting a variational trial solution.

An important contraction is the one eliminating the traffic, which can be performed analytically.
Keeping in mind the symmetry properties of $t_{ij}$ and $j_{ij}$, one finds for
 the optimal value
\beq
{t^*_{ij}}^2=j_{ij}^2+4p_ip_jk_{ij}k_{ji} 
\ee
and, hence, for the rate function \cite{maes08}
\begin{eqnarray}
I(\{p_i\},\{j_{ij}\})&=&I(\{p_i\},\{j_{ij}\},\{t^*_{ij}\})
=\sum_{i<j}\left\{j_{ij} \ln\frac{j_{ij}+\sqrt{j_{ij}^2+4p_ip_jk_{ij}k_{ji}}}{2p_ik_{ij}}
-\sqrt{j_{ij}^2+4p_ip_jk_{ij}k_{ji}}+p_ik_{ij}+ p_jk_{ji}\right\}.
\label{eq:i2}
\end{eqnarray}
In this form, 
the rate function inherits the FT-symmetry (\ref{eq:s-ratio-3})
\beq
I(\{p_i\},\{j_{ij}\})-I(\{p_i\},\{-j_{ij}\})= 
-\sum_{i<j}j_{ij}\ln[p_ik_{ij}/p_jk_{ji}]\equiv -\sigma(\{j_{ij}\}),
\ee
with its time-antisymmetric part given by the 
corresponding entropy production.

\subsection{A universal bound on current fluctuations}
A further contraction of (\ref{eq:i2}) to get rid off the empirical densities can not be performed
analytically. One can, however, get an upper bound on the rate function for the currents
 by replacing
(the unknown optimal) $p_i$ by (the stationary) $p\ss_i$, 
\begin{eqnarray}
I(\{j_{ij}\})&\leq &I(\{p\ss_i\},\{j_{ij}\})
=\sum_{i<j}\left\{j_{ij} \ln\frac{j_{ij}+\sqrt{j_{ij}^2+{t\ss_{ij}}^2-{j\ss_{ij}}^2}}{j\ss_{ij}+t\ss_{ij}}
-\sqrt{j_{ij}^2+{t\ss_{ij}}^2-{j\ss_{ij}}^2}+t^s_{ij}\right\}.\label{eq:i3}
\end{eqnarray}
This upper bound still obeys the FT-type symmetry
\beq
I(\{p\ss_i\},\{j_{ij}\})-I(\{p\ss_i\},\{-j_{ij}\})
= - \sum_{i<j}(\sigma\ss_{ij}/j\ss_{ij})j_{ij}=-\sigma(\{j_{ij}\}),
\ee where
\beq
\sigma\ss_{ij}\equiv j\ss_{ij}\ln[p\ss_ik_{ij}/p\ss_jk_{ji}]=j\ss_{ij}\ln\frac{t\ss_{ij}+j\ss_{ij}}{t\ss_{ij}-j\ss_{ij}}
\ee
is the mean entropy production rate in the link $(ij)$.
 Remarkably, a quadratic function with the same minimum and the same symmetry  provides
a global upper bound on the right hand side of (\ref{eq:i3}) leading to \cite{ging16}
\beq
I(\{j_{ij}\})\leq \sum_{i<j}\frac{\sigma\ss_{ij}(j_{ij}-j\ss_{ij})^2}{{4j\ss_{ij}}^2} .
\ee 
This bound is tight at
$j_{ij}=\pm j\ss_{ij}$ and has, in general,  a larger curvature in the minimum than (\ref{eq:i3}).
Thus the fluctuations of the current  through any link have been shown to be
 larger than a Gaussian
involving the local entropy production.

Finally, to get an upper bound on the rate function for an arbitrary current
$j_\alpha$ with mean value $j\ss_\alpha$, we can choose $j_{ij}=j^s_{ij}j_\alpha/j\ss_\alpha$
leading to
\beq
I(j_\alpha)\leq  \frac{\sigma(j_\alpha-{{j_\alpha}}\ss)^2}{4{{j_\alpha}\ss}^2} .
\label{eq:curr-ldf}
\ee
Hence, the rate function for any current is bounded by a simple
quadratic function whose curvature is determined by the dissipation rate
as first  conjectured in \cite{piet15} and proven along the lines shown here in \cite{ging16}, see also \cite{ging16a}.

\subsection{Bounds on the rate function for empirical densities and traffic}
By applying a similar reasoning, Garrahan has  derived related bounds for non-negative
time-symmetric quantities like  empirical densities and traffic \cite{garr17}.
Rather than entropy production $\sigma$, the overall activity or traffic
$\sum_{ij} p^s_ik_{ij}$ then plays a crucial role.
 
\section{Thermodynamic uncertainty relation: The cost of precision and thermodynamic inference}

\subsection{Formulation}
In significantly larger generality than  for the asymmetric random walk
discussed above in Sect. \ref{sect:arw}, 
the thermodynamic uncertainty relation provides a universal bound on the
precision of any biomolecular process. In a NESS, with each stationary
 current $j_\alpha^s=\sum_{i<j}d^\alpha_{ij}j^s_{ij}$, see (\ref{eq:curr}), 
there is associated a fluctuating "output" $X_\alpha= \sum_{ij} n_{ij}d^\alpha_{ij}$
with mean $\langle X_\alpha\rangle = j\ss_\alpha t$.
From variance and mean of this output in the long-time limit, we define its precision
\beq
\eps^2_\alpha \equiv 
\langle(X_\alpha-j\ss_\alpha t)^2 \rangle/(j\ss_\alpha t)^2 \to 2 D_\alpha/({j\ss_\alpha}^2 t)~~ {\rm for~~ large~~} t, 
\label{eq:prec}
\ee where $D_\alpha$ is the dispersion of the process.
 On the other hand running this process for a time $t$ generates on average
$C=\sigma t$ entropy, which is the thermodynamic cost associated with it.
The thermodynamic uncertainty relation 
\beq
\lim _{t\to \infty} C  \eps_\alpha^2 = 2\sigma D_\alpha/{j\ss_\alpha}^2\geq 2 
\label{eq:unc}
\ee
holds for any Markov process. Thus, precision in the outcome
of any such process requires a minimum cost.

This relation was formulated as a conjecture in \cite{bara15} based on analytical results
for the linear response regime of multicyclic networks and for unicyclic networks
with their only one 
independent current. With the bound (\ref{eq:curr-ldf}) on the rate function, the proof follows trivially using $D_\alpha=1/[2I''(j^s_\alpha)]$ \cite{ging16}. Recent work conjectures it to be true even for a finite time $t$ with $\eps_\alpha(t)$ \cite{piet17}.

\subsection{Thermodynamic inference for a molecular motor}

The thermodynamic uncertainty relation can be used to infer physical properties of
biomolecular systems from the observation of fluctuations even if the underlying
biochemical or enzymatic network is not  (fully) known as we will now illustrate
for a molecular motor running against a constant force $f$ at a mean velocity $v$
with dispersion $D$. 

Any such motor delivers a mean output power $P\out=fv=\sum_{i<j}f j^s_{ij}d_{ij}$,
where $d_{ij}$ denotes the distance the motor steps in a transition $i\to j$ along its
track against the force. The corresponding fluctuating current $j\out$ is a genuine current to which the
uncertainty relation will be applied below. Likewise, this motor is powered by the consumption of ATP leading to a mean input power $P\inp$ that is typicallly not directly accessible.

In a NESS, the entropy production rate, i.e., the rate of wasted free energy,
 can then be written
as
\beq
\sigma = \beta(P\inp-P\out) .
\ee
The thermodynamic efficiency of such a motor $
\eta\equiv  P\out/P\inp
$
fulfills a universal bound set by the thermodynamic uncertainty relation (\ref{eq:unc}) applied to the output current that can be obtained through
a simple algebraic transformations as \cite{piet16b}
\beq
\eta = \frac{P\out}{P\out+\sigma/\beta}=\frac{vf}{vf+\sigma/\beta}\leq \frac{1}{1+v/(Df\beta)} .
\label{eq:eta-mm}
\ee

 The intriguing aspect
of this bound arises from the fact that $v,D$ and $f$ are experimentally accessible
quantities. No knowledge of the underlying network, i.e., of the specific reaction scheme is necessary
for applying this bound. 
There could be idle cycles where ATP is used without
advancing the motor. It is not even necessary to know the free energy difference $\Delta \mu$
associated with the ATP hydrolysis. 
In Fig. \ref{fig:eta-mm}, this bound is evaluated with experimental  data for a kinesin motor reported in \cite{viss99}.

\begin{figure}[htb]
	\centering
	\includegraphics[scale=.7]{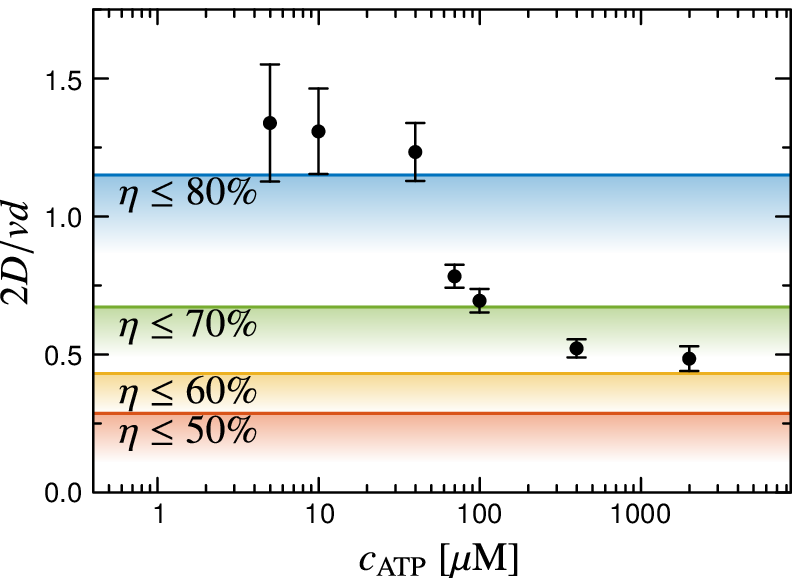}~~~~~~~
	\includegraphics[scale=.7]{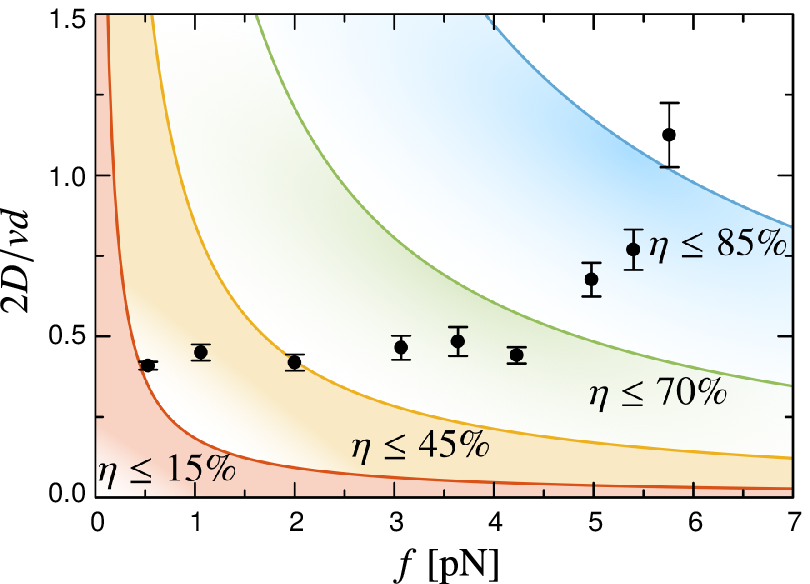}
	\caption{Randomness parameter $r\equiv 2D/vd$  for kinesin as experimentally measured in \cite{viss99} as a function of ATP-concentration
	(for a fixed force $f=3.59$ pN, left panel) and of load force (for fixed $c_{\rm ATP}=2$mM, right panel). The colored area shows the corresponding theoretical bound (\ref{eq:eta-mm}). At, e.g., 2pN load force (right panel), these experimental data imply that 
	this motor thus converts ATP to mechanical power with an efficiency of at most 45 \% for these conditions.}
	\label{fig:eta-mm}
\end{figure}

\section{Topology- and affinity-dependent bounds for thermodynamic inference}

\subsection{Cycles and their affinities}

Cycles and the currents running through them are even better suited  for relating statistical with thermodynamic properties than the currents through individual
links \cite{schn76}.
A cycle $C_a$ is a directed, self-avoiding, closed path of length $N_a$ on the set of states,
see Fig. \ref{fig:cycles}. Its
adjacency matrix $\chi^a_{ij}$ has element 1 if the cycle passes the link $(ij)$ 
in forward direction, -1 if it passes this link in backward direction, and 0 if the
link is not part of this cycle. For a complete set of cycles, all stationary
currents can be expressed as a linear combination of cycle currents
\beq
j\ss_{ij}=\sum_a \chi^a_{ij} j\ss_a.
\ee

In a NESS, after completing any cycle,
the system has returned to its original state and hence all physical 
changes associated with the
cycle have taken place in the surrounding reservoirs. 
The mean entropy production (\ref{eq:sigma}) is time-independent
and becomes a linear combination of cycle currents
\beq
\sigma= j^s_\sigma=  \sum_{i<j}j\ss_{ij} \ln[p_i\ss k_{ij}/p_j\ss k_{ji}]=
\sum_{i<j}j\ss_{ij} \ln[k_{ij}/k_{ji}] =
\sum_a j\ss_aA_a
\label{eq:sigma2}
\ee
where the cycle affinity
\beq
A_a=\sum_{i<j}\chi^a_{ij}\ln[k_{ij}/k_{ji}] = \sum_I n^\gamma_a A^\gamma
\ee
is determined by the ratio of forward and backward rates along a cycle. These cycle affinities are 
(integer) linear combinations of a set of physical affinities $A^\gamma$
that are imposed by the external conditions. Examples for such physical affinities are
$\beta fd$, where $f$ is a force and $d$ a repeat distance on a filament or $\beta\Delta \mu$ of an ATP
hydrolysis. In the example of the ARW from Sect. \ref{sect:arw}, there is the equivalent of only one cycle
with one cycle affinity $A=A\inp-A\out= \beta(\Delta \mu -fd)$.

\begin{figure}[htb]
	\centering
	\includegraphics[scale=.9]{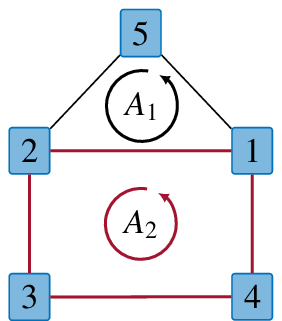}~~~~~~~~~~~
	\includegraphics[scale=.7]{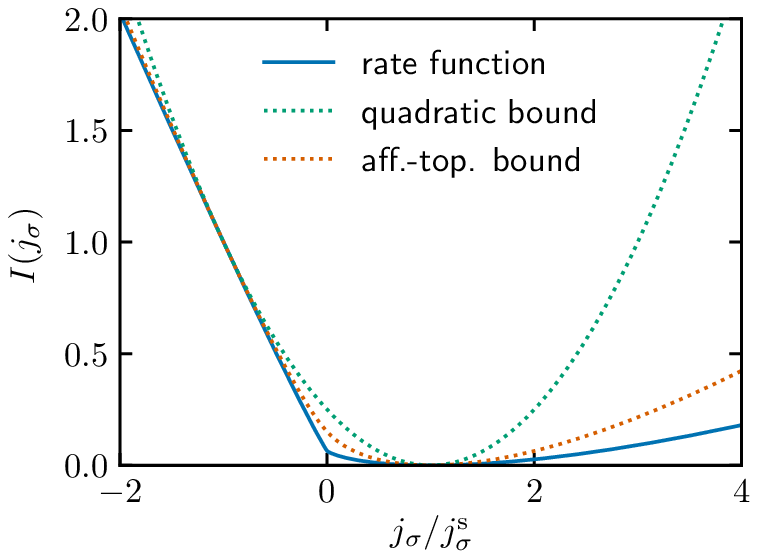}
	\caption{A network with two independent cycles [(1,5,4) and (1,2,3,4)] (left panel). Rate function $I(j_\sigma)$ for the entropy current, its quadratic bound (\ref{eq:curr-ldf}) and the corresponding topology- and affinity-dependent bound (\ref{eq:i4})
	(right panel).
	Transition rates:
	$k_{15}=k_{52}=k_{21}=k_{34}=e^4$, 	 $k_{12}=k_{25}=k_{51}=k_{14}=k_{43}=k_{32}=e^{-4}$, $k_{23}=e^{10}$, $k_{41}=e^6$, leading to the cycle affinities $A_1=A_2=24$ and hence $(A/n)^*=24/4=6$.}
	\label{fig:cycles}
\end{figure}

\subsection{Uniform, unicyclic asymmetric random walk}

For a unicyclic network of $N$ states with uniform rates $k_+,k_-$, hence affinity $A=N\ln(k_+/k_-)$,
and mean current $j\ss=\sigma/A$,
the rate function for the probability current can be obtained
from (\ref{eq:arw-rate}) through contraction or from (\ref{eq:i2}) using the obvious symmetry
$p_i^*=1/N$. In any case, it leads to \cite{lebo99}
\beq
I(j)=I(\xi j^s)=(N/A)\sigma\left[\xi\ln\frac{a\xi+\sqrt{a^2\xi^2+1}}{a+\sqrt{a^2+1}}-\sqrt{\xi^2+1/a^2}+\sqrt{1+1/a^2}\right],
\label{eq:i4}
\ee
with the scaled current
$
\xi\equiv j/j\ss
$
and
$a\equiv \sinh(A/2N) .
$

\subsection{Nonuniform unicyclic and multicylic networks}

For a unicyclic network with non-uniform rates with cycle affinity $A=\sum_{i<j}\ln[k_{ij}/k_{ji}]$, one
 can prove  that (\ref{eq:i4}) provides an upper bound on the rate function for the
 probability current \cite{piet16}. The physical reason is that at fixed affinity and number of states
 uniform rates lead to the smallest fluctuations. 
 
An expansion of (\ref{eq:i4}) around the minimum then implies 
 for the dispersion
coefficient the inequality
\beq
D =\frac{1}{2I''(j^s)}\geq\frac{A}{2N}\frac{{j\ss}^2}{\sigma}\coth(A/2N) ~~~[= (j\ss/2N)\coth(A/2N)],
\label{eq:bound-D}
\ee
which leads for cost and precision to the improved inequality
\beq
C\eps^2=2\sigma D/{j^s}^2\geq (A/N)\coth(A/2N)\geq \max (2,A/N).
\label{eq:bound-unc}
\ee
The first inequality is saturated for uniform rates, the second one close to
equilibrium ($A\ll N$) for the
first choice, and far from it for the second one.

An {\sl a priori} surprising result is that for an arbitrary current $j_a$ 
in a multicyclic network the rate function $I(j_a)$, the  diffusion constant
$D_a$ and the product $C \eps_a^2$ in the refined uncertainty relation are still bounded
by the expressions (\ref{eq:i4}), (\ref{eq:bound-D}, unbracketed) and (\ref{eq:bound-unc}), respectively,
 if one replaces $(A/N)$ by $(A/N)^*$
which is the smallest strictly positive value of $(A_a/ N_a)$ among all cycles
\cite{piet16}, for an example, see Fig. \ref{fig:cycles}. This proof relies on an identification of a suitable choice of
fundamental currents and is somewhat technical. The physics behind it reflects the
fact that the cycle with the smallest $A_a/N_a$
potentially leads to the smallest fluctuations and thus provides a lower bound
on the true fluctuations.

\subsection{Example: Cost and precision of a Brownian clock}

For an example illustrating these concepts, we consider a simple model
for a  thermodynamically consistent clock \cite{bara16}. It consists of
a unicyclic network of $N$ states driven by an affinity $A$ leading to a mean current $j^s$.
A unit of measured time is counted whenever the transition from $N$ to 1 occurs. For consistency,
we have to allow that this transition may occasionally happen in the reversed direction, in which case time has "advanced" a negative unit. After real time $t$,
the clock has measured $X(t)$ units with mean $j^st$ and precision as given by (\ref{eq:prec}).
 The implications of the bound (\ref{eq:bound-unc}) for
the design, precision and cost of such a Brownian clock
can best be illustrated by comparing two clocks using
familiar notions \cite{bara16}. Suppose we want to measure reliably,
say with a precision $\eps=0.01$, a time of one hour with
either a “slow” clock that takes one minute for a revolution or a “fast” clock that takes only one second
implying $\langle X\rangle$ = 60 and 3600 for slow and
fast, respectively. With the cost $C=A\langle X\rangle$, 
the second inequality in (\ref{eq:bound-unc}) 
 implies, first, 
a structural constraint on the minimal number of states
making up the cycle, which is $N_{\min}=167$ and 3 for the slow and the fast clock, 
respectively. The slow clock has to have sufficiently many states within
its cycle to achieve the required precision. Second,
 for a given design, i.e., for a given number of states $N$ in the cycle, the 
 affinity driving the clock has to be at least
$A_{\rm min}= 2 N{\rm arccoth}(\langle X \rangle N \eps^2)
\geq 2/(\langle X\rangle \eps^2)$. For the slow clock,
we get $A_{\rm min} \simeq 333$ and for the fast one,
$A_{\rm min} \simeq 5.55$.
 The overall entropy production associated with measuring one hour
with this precision is bounded by 20000 for both types.
From an energetic point of view, both designs are equivalent.
In a biochemical network the free energy is often provided by ATP hydrolysis, which under physiological conditions liberates approximately 20 $k_BT$ of free energy. 
The universal result $C\eps^2\geq 2$ then implies
that for an uncertainty of $0.01$ the Brownian clock 
requires the consumption of at least
1000 ATP molecules.

The fact that the bound (\ref{eq:bound-unc})
 holds even for the "best" cycle of a
multicyclic network implies that a more intricate "wiring" of the
network cannot improve the inevitable trade-off between precision
and cost. It has been shown, however, that driving such a clock not
by a constant affinity but rather by a modulation of energies and
barriers, i.e. of the transition rates between  the states in a periodic fashion, a given precision
requires no minimal cost \cite{bara16}, see also \cite{rots16}. On the other hand, if one 
includes the thermodynamic cost of providing such a time-periodic variation of parameters,
one is effectively back at the above inequality \cite{bara17a}.

\subsection{Thermodynamic  inference: Fano factor in enzyme kinetics}

These topology- and affinity-dependent bounds can be used as a diagnostic  tool to infer properties of an unknown underlying biochemical
network if the fluctuations of a current can be measured and if some information on the driving
affinity is known. As an example consider an enzyme E that transforms a substrate S into
a product P using hydrolysis of one ATP molecule which liberates $\Delta \mu$ of free energy. 
Suppose in a single-molecule experiment
one measures that after a long time $t$ on average $\langle X\rangle = j\ss t$ product molecules have been generated
with a variance $\langle (X-j\ss t)^2\rangle=2Dt$ which defines the dispersion 
of this current.
With $\sigma= j^s \Delta \mu$, the bound (\ref{eq:bound-D}) implies a bound on the Fano factor \cite{bara15a}
\beq
F\equiv 2D/j\ss\geq (n/N)^* \coth [(\Delta \mu/2)(n/N)^*] ,
\label{eq:fano}
\ee where $(n/N)^*$ is the smallest value for the ratio between number of products $n$ and number of
states $N$ among all cycles.  For a simple Michaelis-Menten scheme with only three states (E$\to$ ES$\to$EP$\to$E, i.e., $N=3, n=1$) the
bound reads $F\geq [\coth(\Delta \mu/6)]/3$. Any measurement that leads to a smaller value
for $F$ implies either that (at least) a fourth state is involved or that the enyzme is able to bind two substrates \cite{bara15a}, see Fig. \ref{fig:fano}.

\def\tt{$\to$}

\begin{figure}[htb]
	\centering
	\includegraphics[scale=1]{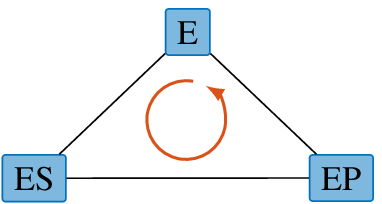}~~~~~
	\includegraphics[scale=1]{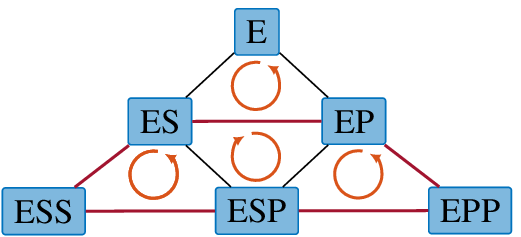}~~~~~
	\includegraphics[scale=.8]{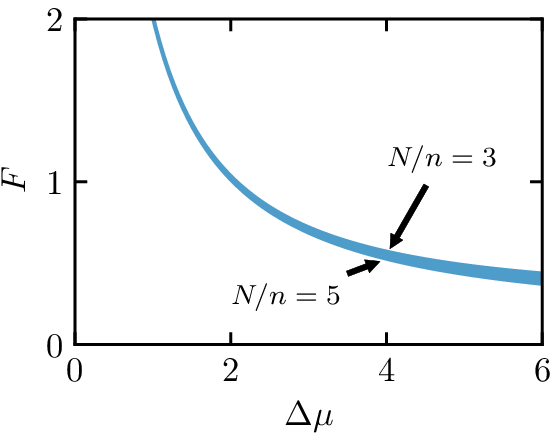}
	\caption{Left panel: Simple Michaelis-Menten scheme for an enzyme E binding a substrate
	S, transforming it into a product P and releasing it. Middle panel:
	 Network for an enzyme that
	can bind two substrates and transform them into products. Right panel:
	Lower bound of the Fano factor (\ref{eq:fano}) as a function of $\Delta \mu$
	for various values for $N/n=3$ and 5, the latter being relevant for
	the 5-state cycle ES\tt ESS\tt ESP\tt EPP\tt EP\tt ES. The colored region in between is allowed for
	the scheme in the middle and excluded for the simple scheme from the left.}
	\label{fig:fano}
\end{figure}

\section{Concluding remark}

The basic principles of stochastic thermodynamics as recalled here are by
now firmly established. Whenever a driven system is connected to a heat bath
and a set of slow variables can be identified whose dynamics is well-separated
from that of the unobserved fast degrees of freedom, thermodynamic
quantities like work, heat and entropy production
can be identified. Their distributions obey universal  
fluctuation relations that have been measured computationally and experimentally
in many systems. As the second part of these lectures is supposed
to 
demonstrate, we are arguably now entering a second stage where inequalities
like the thermodynamic uncertainty relation, which have been derived by
following the consistency conditions imposed by
stochastic thermodynamics, are used to infer hidden properties of a system,
a strategy that could be called  "thermodynamic inference" \cite{alem15}. Quite likely, many exciting insights into the
operation of small machines will be unravelled as these concepts are combined
with single molecule experiments. These insights will not be confined to the
isothermal realm of biomolecular systems. Analogous progress has indeed been made
for heat engines operating between baths of two different temperatures
as the identification of a universal trade-off between power, efficiency and constancy of operation
paradigmatically shows \cite{piet17a}.
\vskip 5mm
{\bf Acknowledgments:}
The results developed in the second part of these lectures have originally
been obtained with Andre C. Barato and Patrick Pietzonka. I thank both for an enjoyable ongoing collaboration and the latter for a careful reading of these notes. I am grateful to Marco Baiesi, Alberto Rosso and Thomas Speck
for the opportunity to present these lectures at this summer school. The help
of Patrick Pietzonka and Matthias Uhl with preparing the figures is gratefully acknowledged.

\section*{References}


\end{document}